# Verhulst Equation and the Universal Pattern for the Global Population Growth

Agata Angelika Sojecka[1] and Aleksandra Drozd-Rzoska[2]

[1]University of Economics, Department of Marketing, ul. 1 Maja 50, 40-257 Katowice, Poland

[2]Institute of High Pressure Physics Polish Academy of Sciences, ul. Sokołowska 29/37, 01-142 Warsaw, Poland

[1]ORCID: 0000-0003-4121-31521; e-mail: agatka.angelika@gmail.com

[2]ORCID: 0000-0001-8510-2388; e-mail: arzoska@unipress.waw.pl

**Abstract**

The global population growth [$P(t)$] from 10,000 BC to 2023 is discussed within the Verhulst scaling equation and its extensions framework. The analysis focuses on the per capita global population rate coefficient $G_P(P) = (dP(t)/P(t))/dt = dlnP(t)/dt$, which reveals two linear domains: from ~ 700 CE till 1966 and from ~ 1966 till 2023. Such a pattern can be considered a universal reference for reliable scaling relations describing $P(t)$ changes. It is also the distortions-sensitive validity test indicating domains of their applicability and yielding an optimal value of parameters. For models recalling the Verhulst equation, single pairs of growth rate and system capacity coefficients $(r, s)$ should describe global population rise in the mentioned periods. However, the Verhulst relation with such effective parameters does not describe $P(t)$ changes. Notable is the new way of data preparation, based on collecting data from various sources and their numerical filtering, to obtain a 'smooth' set of optimal values, enabling the derivative-based analysis. The analysis reveals links between $G_P(P)$ changes and some historical and prehistorical references, influencing the global scale.

1. **INTRODUCTION**

The Anthropocene epoch began 10,000 years ago due to global warming terminating the last ice age. Approximately 1 million people were living on Earth at that time. Almost 12,000 years later, in 1800, the global population increased 1,000 times, reaching 1 billion [1,2]. Only 125 years were needed to increase by another billion. In November 2011, the global population reached 7 billion, and 11 years later, 8 billion [3].

In the 21$^{st}$ century, mobile phones and online information exchange systems supported by artificial intelligence are omnipotent. Industrial production based on global supply chains has become the norm. Once the ongoing process of developing and implementing hypersonic transport terminates, travel between the most distant places on Earth will be reduced to a few hours. 'Globalization,' referring to interactive human populations in the spatially constrained system of the Earth, is becoming a fact. However, this '*Brave New World*' [4] is threatened by the collapse of the social and political ordering, if not the civilization. One can recall fast-spreading pandemics, such as the sequence of Covid mutations, the enormous anthropic Climate & Global Warming crisis, the Energy Crisis, migration waves, and political unrest, leading even to wars. The latter is often associated with global-scale egoistic targets of some persons, states, organizations,.. with no regard for climatic and ecological threats that even threaten human survival.

It might seem that today's times, driven by extraordinary technological innovations and grand problems and challenges, are exceptional. However, people living in England or Scotland at the beginning of the 19$^{th}$ century, when the 1$^{st}$ Industrial Revolution was becoming omnipotent, could have had similar feelings. The Steam Age innovations had been widely implemented, yielding previously unimaginable technological solutions and leading to turbulent political and socio-economic changes. Developing industry-driven cities became overcrowded, noisy, shrouded in choking smoke, and with dramatically polluted rivers [5]. In 21$^{st}$ century, the 4$^{th}$ Industrial Revolution [6,7] times, challenges and problems are similar but at a truly global level.

Consequently, a model view into past population trends and forecasting future changes is essential for global insight, planning, and governance trials. Many national and international agencies and independent researchers focus on modeling the evolution of global population growth. Nevertheless, the problem remains puzzling, as shown by the scatter of the population forecasts ranging between 6.3 and 14.5 billion for the relatively close period 2050-2100 [8]. There are two main cognitive paths for modeling global population changes.

The first focuses on searching for scaling equations describing long-range population changes, which can validate extrapolations for the nearest future. This direction was initiated by the pioneering works of Malthus (1798) [9] and Verhulst (1838) [10]. The latter directly introduced the influence of the systems on population changes, taking into account available resources or, more generally, the system's carrying capacity. Since then, many other scaling equations for modeling global population growth appeared [11-21 and refs therein]. However, Malthus and Verhulst's model relations have remained a primary reference [22-29].

The target of the second direction is not a general 'scaling model/equation' describing changes in the global population. It aims to define/re-create the global population value, taking as basic reference elements different geographical regions, social groups, changes in education, multiple aspects of social interactions - especially regarding the role of women, local migration issues, education, birth/death ratio, .... Such data are analyzed statistically in frames of models developed in management, bio-evolution, or socio-economics sciences [21,30-37], which have shown their effectiveness for various problems from the scale of states to companies and corporations, but also in various populations-related issues in medicine, complex bio-systems, … [21]. For this cognitive path, links between different factors are essential, often in feedback interactions, which, in practice, weightings based on expert opinions must be supported.

This raises the question of subjective arbitrariness and reliable error estimations.

For sub-planetary systems, such as states, regions, and continents...., populations develop in open systems, with interactions and inter-flows with the surroundings. The global population develops in a closed system, defined by Earth's space limits and carrying capacity. Local changes and trends are averaged and compensated globally; only some can be significant for this scale. A subtle insight into past global population changes is essential to recognize such situations, but this remains challenging.

The first inspiration for this report was a recent paper by Lehman et al. [38], which combines the above paths of global population studies and considers the Verhulst scaling equation within the concept proposed by Pearl and Reed [39,40] by introducing multidimensional coefficients describing the increase rate and the carrying capacity parameter. This report critically discusses the Verhulst-type approaches, considering the per capital relative changes of the population growth as the distortions–sensitive validation test. The unique patterns of its changes can be the universal check-in reference for any scaling equation for global population changes. The analysis also enabled subtle insights into global population changes, showing historical and prehistoric references.

**2. VERHULST SCALING EQUATION and ITS EXTENSION by LEHMAN et al.**

The turn of the 18$^{th}$ and 19$^{th}$ centuries was associated with the rising wave of the 1$^{st}$ Industrial Revolution, especially in England. Massive industrial centers exploring breakthrough Steam Age [5] technological innovations appeared. They were overcrowded, full of hope for a new life, but also enormous poverty and social unrest [5]. In these times, the Scientific Method has already become a leading cognitive method supporting the innovations-driven Industrial Revolution. It was primarily due to Isaac Newton legacy, ranging from physics and mathematics to economics [41]. Newton showed the ultimate importance of empirical verification and adequate descriptions of the laws of nature using functional scaling relations, including the differential analysis he introduced. His legacy is also the general belief that apparently separate phenomena can be described by universal laws of nature.. The unified description of the motion of an apple falling from a tree and of planets and comets 'in the sky'' was a crucial example of Newton's grand success [41].

These inspirations were declared by Robert Malthus, who formulated the first and still important model equation for describing population changes $P(t)$ [9]:

$$\frac{dP(t)}{dt} = rP(t) \quad \Rightarrow \quad \frac{1}{P(t)}\frac{dP(t)}{dt} = r = const \tag{1}$$

$$P(t) = P_0 exp(rt) \quad \Rightarrow \quad lnP(t) = lnP_0 + rt \tag{2}$$

where time $t$ refers to the onset time $t_0$ matched to the reference population prefactor $P_0$; the Malthus growth rate coefficient $r = const$.

Equation (1) shows differential equations matched to the Malthus Equation (2). The latter can be validated by the linear behavior for the semi-log plot of $P(t)$ data, in the right-hand part of Eq. (2).

In 1838, Pierre François Verhulst pointed out that the features of the system in which a given population develops may be of fundamental importance [10]. He focused on available resources, originally food, which led him to the differential equation [10,39,40]:

$$\frac{dP(t)}{dt} = rP + sP^2 \quad \Rightarrow \quad \frac{dP(t)}{dt} = rP(t)(1 - sN) = rP(t)\left(1 - \frac{N}{K}\right) \tag{3}$$

where the coefficient $s$ describes available resources and $K = 1/s$ is called the *carrying capacity*: it describes the maximal population that available resources can support.

The left part of Eq. (3) refers to the original work of Verhulst [10], and the right one, with the carrying capacity $K$, was introduced by Pearl and Reed [39,40]. The integration of Eq. (3) yields the following scaling equation [10,39,40]:

$$P(t) = \frac{K}{1+CKexp(-rt)} = \frac{1}{1/K+Cexp(-rt)} \quad \Rightarrow \quad (K \to \infty) \Rightarrow P(t) = P_0 exp(rt) \tag{4}$$

where $C = 1/P_0 - (1/K)$

The right-hand part of Eq. (4) shows the transformation to Malthusian Eq. (2).

In frames of the Verhults relation, one can consider two basic cases. For systems with renewable resources and carrying capacity, the metric $K(t) = const$, despite passing time and population changes. In such a case, after the initial Malthusian growth, a transition to the stationary phase with a constant, equilibrium, population $P(t) = const$ occurs. For systems with non-renewable resources & capacity, the stationary phase is relatively short, and population decline occurs, also of the Matusian type but with a negative coefficient $r$. Finally, the population may disappear due to resource exhaustion. Such a picture constitutes a standard in microbiological tests for populations of bacteria or yeast in a container isolated from the surroundings and a given and unreplenished amount of food (like sugar) [42-47]. Recently, such a pattern has also been shown for human population changes on Easter Island (Rapa Nui), the island on the Pacific Ocean, located well remote from other islands and the South America mainland. Recently, it was also shown for industrial cities created by a dominant industry [42]. It is the case of Detroit (IL, USA), which is associated with the automobile industry, and Bytom (Silesia, Poland), a former coal mining center [42].

Malthus and Verhulst modeling is the reference tool for modeling population changes in many branches of science, from microbiology [43-46] and food technology [47,48] to the spread of epidemic outbreaks [49], growth of some animals and plant populations [43,46,50] to some problems in economy and management [51-53], and physics [54-56].

There is also a third, rarely discussed option of population changes resulting from the Verhulst scaling relations (Eqs. (3) and (4)), especially for isolated (closed) systems with terminated resources and carrying capacity. One can consider a relative resource increase due to a significant reduction in population requirements. In the language of physics, it can be described as *spontaneous self-adaptation of complex active matter population* to the system's constraints. For instance, for populations of animals, it may mean the ability to make fast, spontaneous, adaptive evolutionary changes.

To illustrate the '3rd route', one can recall the case of pygmy mammoths [57]. Near 10,000 BC, rising ocean levels cut off mammoths from the west coast of North America on Channel Island [57,58]. The last of them lived only 4,000 years ago. The evolution caused their height to be only 1.7-2 m, and their weight was even 10x less than that of basic Columbian mammoths. Such reduction led to the new equilibrium, increasing the number if available resources and the available space. A new sustainability between the system and the population was reached, allowing for its more prolonged survival. It has been suggested that the ultimate disappearance of the pygmy mammoth was mainly due to genetic degeneration, i.e. 'internal' population problems.

For the global human population evolving within the Earth's spatial, resource, and ecological capacity, the 3rd path can mean a sustainable civilization with rational energy consumption, minimal environmental harm, and using only durable and long-lived products. Sustainable civilization is a recommended development path to avoid the global-scale threats mentioned above.

A question arises if such a 'sustainable society' strategy has already appeared in the past and reached a 'population success'?

For the authors, origins of Slavic tribes in early Middle Ages are worth considering as a possible example. Pre-Slavic tribes appeared in Central Europe 'suddenly' between 5th and 7th centuries CE. It was a time of climatic breakdown, the peak of which was the so-called Justinian winter, associated with the temperature in Europe, and perhaps globally, dropping by as much as 1-2 K. In Central Europe, winters became long and very cold [59]. It led to essential vegetation and crop yield problems for farming communities. Such conditions were one of the motivations for the great migrations of Germanic tribes from Central Europe to the Roman

Empire located in a more favorable climate. Finally, it led to the fall of the Western Roman Empire, and the long-term problems of its eastern part associated with Constantinople [60]. Suddenly, in Central Europe's 'abandoned' areas, traces of small communities with a surprisingly primitive, or perhaps only poor, way of life appeared. They are associated with pre-Slavic tribes, whose original habitats are related to unspecified locations in 'deep' Eastern Europe [61,62]. However, recent genetic research has shown that the ancestors of the proto-Slavics lived in central Europe at least 500 years ago [63], probably coexisting peacefully with Germanic tribes. During the 'climate catastrophe' times, Germanic tribes chose migration to solve the problem, which terminated with the conquest of the Western Roman Empire. A part of the population, probably very closely related to agricultural life, seems to have remained in Central Europe. The symbol of their 'primitivism' was living in dugouts, previously unknown in this area [61,62]. But such dugouts is also the most effective ways to survive under extreme conditions. For the authors, this situation can be seen as a transition to a "sustainable society", adapted to the conditions of the climate crisis.. It could also be a significant formative period for Slavic tribes and the source of their enormous success in the 8$^{th}$ and 9$^{th}$ centuries [61], as the climate warmed rapidly and available resources and opportunities increased.

In the Anthropocene, strong and permanent super-Malthusian global population growth occurs. It is expressed via nonlinear changes in the semi-log plot $lnP(t)$ or $log_{10}P(t)$ vs. $t$ [64]. Such a pattern is beyond the canonic Malthus behavior and Verhulst behavior discussed above (Eqs. (1) – (4)). Nevertheless, a century ago, Pearl and Reed [39,40] suggested that human population growth may follow a sequence of Verhulst scaling equations coupled with a succession of carrying capacities for which the transition to the next one occurs well before the end of the previous one is approached. The population growth pattern may pass through successive Verhulst-type steps without any distinctive manifestation of the Verhulst plateau. Such an analysis carried out in 1920 made it possible to describe population changes until 1930 [39,40]. In the subsequent decades, the growth of the USA population was significantly greater, which can be associated with increasing immigration and the fact that, in a given aspect, the USA is an open system, contrary to the global population. In 1928, Volterra [65] suggested linking mentioned carrying capacity crossovers can be associated with overcoming subsequent significant bio-ecological barriers. Very recently, Lehman et al. [38] have developed these concepts by considering global population growth in terms of three successive bio-ecological levels: (1) predator interactions, (2) prey interactions, and (3) intraspecific interactions.: (*a*) interactions with predators, (*b*) interactions with prey, and (*c*) interactions within. Global population changes by each 'level' are governed by level-dependent coefficients $(r_i, s_i)$, $i = 1,2,3$. The values of these coefficients may even change signs for subsequent stages of human development in the Anthropocene. They are related to (*i*) primordial, (*ii*) tools, files, and specializations, (*iii*) agriculture, and (*iv*) controlled fertility. The following form of the reference Verhulst differential equation was first considered [38]:

$$\frac{1}{P(t)}\frac{dP(t)}{dt} = r_i \pm s_i P(t) \tag{5}$$

where coefficients $r_i, s_i = const$, for subsequent time domains differently subjected to time-varying bio-/eco- factors mentioned above, and the sign'$\pm$' stresses that in ref. [38] both positive and negative coefficients $s_i$ were considered.

Subsequently, global population data from 10,000 BC to 2010, were analyzed using a sequence of $(r, s)$ coefficients and the following form of the Velhurst equation [38]:

$$P(t) = \sum \frac{1}{\left(P_0 \mp (s_i/r_i)e^{-r_i t} \pm (s_i/r_i)\right)} \tag{6}$$

In Eqs. (4) and (5), we introduced the symbols '$\pm$' and '$\mp$' to emphasize that both $s > 0$ and $s < 0$ were considered for subsequent population development stages [38]. One of the justifications for such behavior was linear changes in the per capita growth of the population $G_P(N) = (1/P)(\Delta P/\Delta t)$ plotted against the population itself, with different signs of slopes (i.e. $s$ parameter) [38]:

$$G_P(N) = \frac{1}{P}\frac{\Delta P}{\Delta t} = r \pm sN \qquad (7)$$

where the values of $G_P$ were calculated using the differences between $\Delta N_i$ populations for successive time steps $\Delta t_i$.

Recalling results presented in Fig.3 of ref. [38] one that for the period $10,000\ BCE \ll 1970 \pm 5$ is associated with coefficients $r_1 > 0$, $s_1 >$ and the period $\sim 1970 < t < 2010$ with coefficient $r_2 > 0$, $s_2 < 0$ for. Using results presented in Fig. 3, ref. [38] one can estimate the crossover between these domains at $P(t = 1970 \pm 5) = 3.4 \pm 0.2\ billion$. The authors' of the given report noted that the mentioned behavior extends up to 2023, with $\sim 8.05\ billion$ population. Following results presented in ref. [38], one can estimate the crossover the crossover $G_P(P) > 0 \to 11 \pm 0.5\ billion \to G_P(P) < 0$. It can be considered a hypothetical future population maximum. In ref. [38] a set of 36 values of $(r_i, s_i)$ parameters for subsequent time domains were considered. The model considered in ref. [38] based on 98 global population data from ref. [38], covering nearly 12 millennia. Half of the data concerned the period after 1970. The graphical extension to subsequent decades of the 21st century was plotted.

### 3. METHODOLOGY

This report explores the new way of data preparation based on collecting global population data from various sources and their numerical filtering using the protocol introduced by one of the authors (ADR) in material engineering and glass transition physics studies [66-69]. It enables finding optimal evolution paths in a set of inherently scattered 'noise-like' data. It employs the Savitzky-Golay principle with the support of Origin and Mathematica software. They explore global population data from ref. [70-77]. As the analysis result, a 'smooth' set of 193 population data, from 10,000 BC to 2023, has been obtained and presented in the Appendix.

Such data preparation enabled their derivative analysis, avoiding problems with estimations of the global population, which differ even by 100 % simultaneously. The above enabled the linearized distortions-sensitive and derivative-based test of the global population, for which the linear domain indicates the domain in which a scaling equation can be applied, and the linear regression yields optimal values of relevant parameters with well-defined errors [66-69].

## 4. RESULTS AND DISCUSSION

Figure 1 shows global population changes from Anthropocene (10,000 BC) onset to 2023, based on data prepared in this report and given in the Appendix. The inset focuses on the ongoing Industrial Revolutions [6,7] times. The arrows indicate emerging characteristic changes in the evolution of the global population.

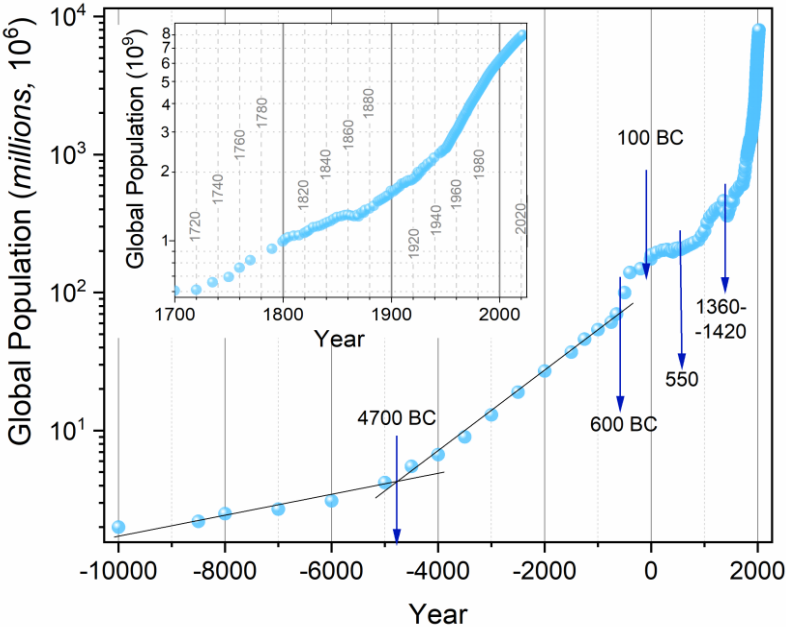

**Figure 1** The plot of global population change, in a semi-logarithmic scale, from 10,000 BC to 2023, based on the data given in the Appendix. The inset focuses on the Industrial Revolutions [6,7] epoch. The arrows indicate some characteristic dates/periods that appear in the plot.

For almost 10,000 years, up to ~ 600 BC, which can be correlated with the definitive end of the Bronze Age, or the development of great civilizations in the Mediterranean area and China [78], global population evolution can be well portrayed by the Malthusian relation (Eq. 2), shown by the linear behavior for the semi-log plot in Fig. 1. However, there is a significant change in the slope of such behavior around 4700 BC, which may be related to the acceleration of population growth: the Malthus rate coefficient increased 4.6x after 4700 BC. The population crossover correlates with the transition from the Palaeolithic to the Neolithic times [78]. Between 100 BC and 500 AD, a plateau in global population changes appears. It remains constant at 190-200 million level. This period correlates with the Roman Empire times [79]. Its population is estimated to reach 40 million, but even 70 million has been recently estimated at its peak [80]. It means that the Empire encompassed between $1/5$ to $1/3$ of the global population. The enormous success and the fall of the Roman Empire has remained the subject of research and fascination for generations of historians [79-81]. We want to draw attention to only one important element, important in frames of the Verhust model discussion.

In Roman Empire times, slavery was a 'social norm'. But enslaved people have in the Empire an additional meaning; namely, they were also the crucial 'energy resource' that drove the economic 'machine' and were exploited on the extreme 'global' scale. The enslaved built ompressive buildings, aqueducts, channels, and tunnels that remain symbols of Roman Empire times also nowadays. For instance, in Rio Tinto (Iberia) giant silver mines, between 20 and 50 thousand enslaved people worked. The great historian Pliny (*Gaius Plinius Secundus*) noted that each could survive between 6 months and 2 years [81]. Using the modern language, Roman 'managers' had such 'energy consumption' and 'new supplies' in 'business plans'. Terrifying. The imperial economy required a constant influx of 'human energy' - enslaved people. Wars and expeditions into 'barbarian' territories to obtain huge amounts of slaves ('human energy') were necessary for the economic survival of the Empire. For the authors, it may be an important factor in the anomalous stabilization of global population changes at these times. However, the Empire was weakening, and the 'barbarians' were getting stronger. New 'human energy supplies' to fuel the imperial economy were becoming impossible. Following the Verhulst's model approach, the lack of a major resource must lead to the decline and collapse.

Figure 1 also shows the strong impact of the Black Death epidemic that devastated Asia and Europe in the 14$^{th}$ and 15$^{th}$ centuries, leading to a significant decrease in world population [82].

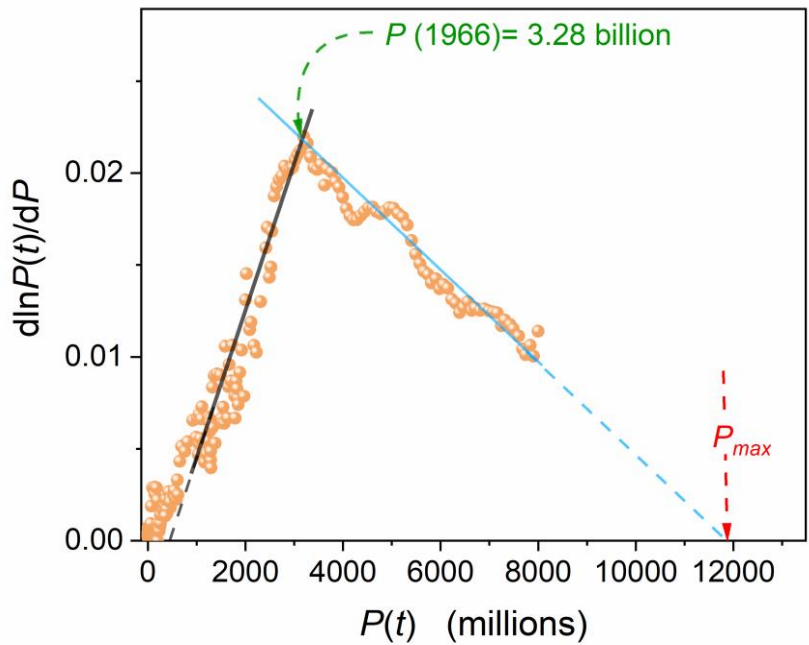

**Figure 2** The per capita growth of the relative world population growth $G_P(P)$ determined by the derivative analysis defined by Eq. (8): It is based on data shown in Figure 1 and collected in the Appendix.

When discussing the global population and its relation to the extended Verhults- type scaling (Eqs. 3,4 & 6) one should consider the extension of Eq. (7) for per capita population growth to the case of 'smooth' population data, where the derivative analysis is possible:

$$G_P(P) = \lim \left[\frac{1}{P(t)}\frac{\Delta P}{\Delta t}\right]_{\Delta t \to 0}^{\Delta N \to 0} \quad \Rightarrow \quad G_P(P) = \frac{dP(t)/P(t)}{dt} = \frac{d\ln P(t)}{dt} = r \pm sP(t) \quad (8)$$

Figure 2 shows the derivative analysis for the global population data shown in Figure 1 considered in frames of Eq. (8). The behavior obtained reasonably agrees with that presented in Fig. 3 of ref. [38] by Lehmann et al., based on Eq. (7). Figure 2 explicitly shows two linear domains with a crossover near ~ 1966. A similar behavior, with a crossover ~1970 can be concluded from results presented in ref. [38], with the first linear domain extending down to 10,000 BC. The logarithmic scale can minimize the impact of changing tested magnitudes (time, population), by decades which can 'hide' some 'subtle' features. Such results are presented in Figure 3. The explicit linear pattern in $G_P(P)$ changes occur only after ~700 AD and continued until the crossover at 1966-1970. Notable, that this trend began at the times of the King of Franks and Longobards and the Roman Emperor *Charles the Great*, *Charlemagne*, nowadays considered the 'father' of modern Europe [83].

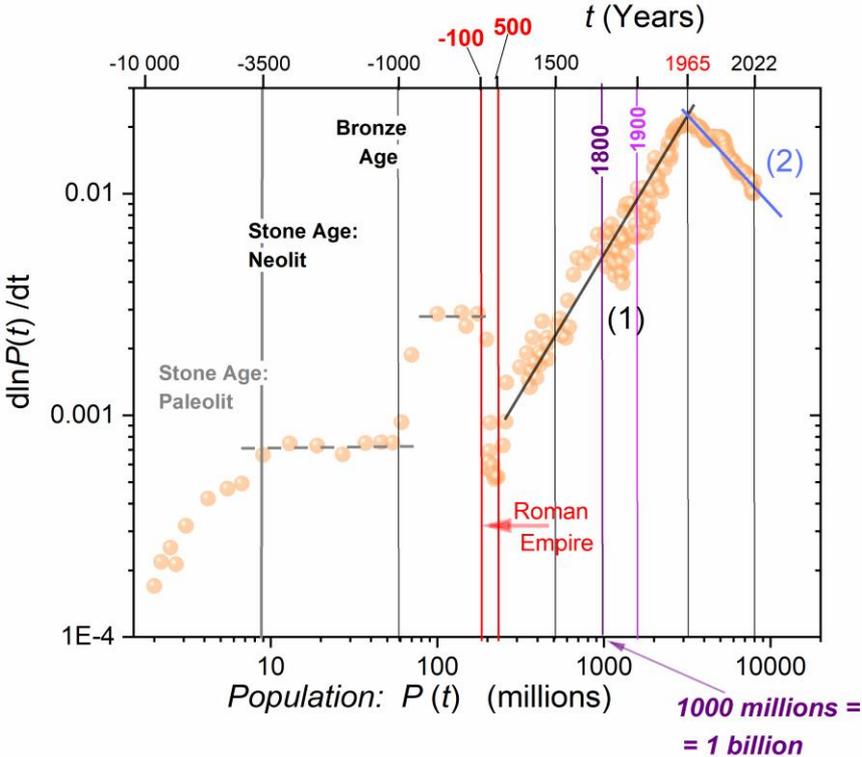

**Figure 3** The log-log scale presentation of the per capita growth of the global population $G_P(P)$, based on the results presented in Fig. 3. Related historical domains are indicated.

Figures 2 and 3 and Eq. (8) can also be considered a subtle, distortion-sensitive validation tool for scaling relations describing global population $P(t)$ changes. Following Eqs. (7) and (8), Figs. 2 and 3, and results presented in ref. [38], two dominant linear domains are related to the Verhulst model Eq. (2). Table I contains parameters describing these domains, determined via the linear regression fit for results presented in Figs. 2 and 3.

Equation (8) can also be considered as the linearized distortion-sensitive and derivative-based test of the Verhulst relation: linear domains indicate regions of its applicability, and the linear regression analysis determines optimal values of $(r, s)$ parameters. It allows the reduction of adjusted parameters in the final $P(t)$ fitting solely to the onset prefactor $P_0$. Such linearized distortions-sensitive analysis proved its effectiveness and reliability in similar tests carried out in physical and material engineering studies in glass forming and so-called critical systems [66-69].

**Table I** Values of the parameter characterizing the linear domains for the per capita population growth rate $G_P(P)$, defined by Eq. (8) and shown in Figure 3.

| time period | population range | $r$ parameter | $s$ parameter | $K = 1/s$ parameter |
|---|---|---|---|---|
| 700 CE - 1967 | 1 million – – 3.3 billion | $\begin{pmatrix} -0.49 \\ \pm 0.05 \end{pmatrix} \times 10^{-2}$ | $\begin{pmatrix} 8.2 \\ \pm 0.1 \end{pmatrix} \times 10^{-6}$ | $0.12 \times 10^6$ |
| 1967 - 2023 | 3.3 billion – – 8.1 billion | $\begin{pmatrix} 2.8 \\ \pm 0.2 \end{pmatrix} \times 10^{-2}$ | $\begin{pmatrix} -2.25 \\ \pm 0.05 \end{pmatrix} \times 10^{-6}$ | $-0.89 \times 10^6$ |

Single pairs of $(r, s)$ parameter for each mentioned domain, given in Table I, should led to the portrayal of $P(t)$ changes after the substitution to the Verhulst scaling relation (Eqs. (4), (6)). However, it does not yield any portrayal of the global population changes in the mentioned domains (see Figs. 2 and 3). A different result could not be expected, considering that in ref. [38] 36 pairs of values $(r_i, s_i)$ was needed for parameterization via the 'extended' Verhults Eq. (6) However, such an application is incompatible with the above discussion, which explicitly shows that single pairs of $(r_i, s_i)$ parameters in the Verhulst dependence are allowed for population-related linear domains (1) and (2) in Figure 3.

## 5. CONCLUSIONS

The multi-parameter application of the Verhulst relation proposed by Pearl & Reed to describe population changes, has been excellently developed in a recent paper by Lehmann et al. [38], in an interesting and convincing model concept. They discussed global population changes $P(t)$ from 10,000 BC till 2010, and also showing the preliminary result for per capita relative global population growth $G_P(P)$. To test the latter magnitude, the author prepares the new set of global population data based on numerical filtering of data from different sources to obtain a 'smooth' set enabling the derivative-based analysis. The derivative-based analysis of $G_P(P)$ confirmed the linear behavior predicted by Eq. (8), split into two domains: (1) ~ 700 CE – 1966, and (2) 1966-2023. It means that single sets of optimal $(r, s)$ parameters, whose values are given in Table I, should describe $P(t)$ changes when substituted to the Verhulst Eq. (4). Unfortunately, this leads to no parameterization. It can be considered a negative validation test of the Verhulst scaling application for describing global population changes. However, the authors want to draw attention to the mentioned above '3r$^d$-way path' discussion for the Verhulst relation, defined as '*spontaneous self-adaptation of complex 'active population*'. This means progressive, adaptive changes that change the population itself and, consequently, the system's carrying capacity associated with it. Perhaps it is necessary to introduce into Verhulst's account a new factor determining population changes in the context of civilization changes. The changes seem to be well reflected by changes in carrying capacity, in this respect, the results of ref. [38] can be an important guide for further research.

However, it is worth noting that changes per capita relative to global population growth $G_P(P)$ defined by Eq. (8), confirmed experimentally in Figures 2 and 3, are a universal characterization of changes in the global population $P(t)$, it is not related exclusively to the Verhulst relation. It is a universal feature of global population changes and, simultaneously, a reference validation test for each model scaling equation $P(t)$ evolution. The global population is developing in a closed planetary system of the Earth, limited in space, limited in the amount of available resources, and with limited environmental and ecological carrying capacity, which is now particularly felt globally. The discussion in this work shows that distortions-sensitive and derivative-based analysis, of which $G_P(P)$ magnitude is an example, and it also indicates several coincidences between the characteristics of global population changes and historical and prehistoric references, indicating which of them had global significance.

**ACKNOWLEDGEMENT:** The work by was supported by the National Center for Science (NCN, Poland), grant ref. 2022/45/B/ST5/04.

**Authors contribution:** The authors equally contributed to the report.

**Data Availability Statement:** Data are given in the Appendix

**Conflicts of Interest:** The authors declare no conflicts of interest concerning the research.

**APPENDIX**

Global population data explored in the given report, and obtained using reference data from refs. [70-77], further subjected to numerical filtering based on the Savitzky-Golay method, according to the protocol developed and tested in refs. [66-69] to obtain a smooth, analytic set of data enabling the derivative analysis. For the 'Years' column the sign '-' is related to BC times.

| Year | Population (milion) | Year | Population (milion) | Year | Population (milion) | Year | Population (milion) |
|---|---|---|---|---|---|---|---|
| -10000 | 2 | 100 | 195 | 1790 | 920 | 1913 | 1793 |
| -8500 | 2.2 | 200 | 202 | 1800 | 990 | 1914 | 1812 |
| -8000 | 2.5 | 250 | 203 | 1802 | 1015 | 1916 | 1837 |
| -7000 | 2.7 | 300 | 205 | 1805 | 1035 | 1918 | 1832 |
| -6000 | 3.1 | 360 | 200 | 1810 | 1050 | 1920 | 1852 |
| -5000 | 4.2 | 380 | 199 | 1815 | 1055 | 1922 | 1887 |
| -4500 | 5.5 | 410 | 197 | 1820 | 1080 | 1924 | 1921 |
| -4000 | 6.7 | 450 | 208 | 1822 | 1094 | 1925 | 2000 |
| -3500 | 9 | 500 | 210 | 1824 | 1109 | 1926 | 1972 |
| -3000 | 13 | 550 | 205 | 1828 | 1144 | 1928 | 2021 |
| -10000 | 2 | 600 | 210 | 1832 | 1154 | 1930 | 2090 |
| -8500 | 2.2 | 650 | 215 | 1835 | 1164 | 1932 | 2116 |
| -8000 | 2.5 | 700 | 220 | 1838 | 1184 | 1934 | 2175 |
| -7000 | 2.7 | 750 | 225 | 1841 | 1203 | 1937 | 2224.5 |
| -6000 | 3.1 | 800 | 230 | 1844 | 1218 | 1940 | 2313 |
| -2500 | 19 | 890 | 240 | 1847 | 1243 | 1945 | 2417 |
| -2000 | 27 | 950 | 260 | 1850 | 1268 | 1947 | 2449 |
| -1500 | 37 | 1000 | 280 | 1854 | 1278 | 1948 | 2493 |
| -1250 | 46 | 1050 | 320 | 1857 | 1292 | 1950 | 2526 |
| -1000 | 54 | 1100 | 353 | 1860 | 1297 | 1951 | 2543.1 |
| -10000 | 2 | 1150 | 370 | 1863 | 1287 | 1952 | 2590.3 |
| -8500 | 2.2 | 1200 | 395 | 1866 | 1278 | 1953 | 2640.3 |
| -8000 | 2.5 | 1250 | 409 | 1869 | 1298 | 1954 | 2691.9 |
| -7000 | 2.7 | 1300 | 415 | 1870 | 1276 | 1955 | 2746.07 |
| -6000 | 3.1 | 1330 | 425 | 1873 | 1327 | 1956 | 2801 |
| -5000 | 4.2 | 1360 | 464 | 1875 | 1325 | 1957 | 2857.87 |
| -4500 | 5.5 | 1380 | 385 | 1876 | 1362 | 1958 | 2916.11 |
| -4000 | 6.7 | 1425 | 360 | 1880 | 1381 | 1959 | 2970.29 |
| -3500 | 9 | 1450 | 380 | 1884 | 1421 | 1960 | 3019.233 |
| -3000 | 13 | 1475 | 410 | 1888 | 1485 | 1961 | 3068.37 |
| -2500 | 19 | 1500 | 457 | 1890 | 1506 | 1962 | 3126.69 |
| -2000 | 27 | 1540 | 465 | 1892 | 1525 | 1963 | 3195.78 |
| -1500 | 37 | 1575 | 530 | 1894 | 1545 | 1964 | 3267.21 |
| -1250 | 46 | 1600 | 544 | 1896 | 1570 | 1965 | 3337.11 |
| -1000 | 54 | 1620 | 556 | 1898 | 1585 | 1966 | 3406.42 |

| -750 | 61  | 1660 | 590 | 1900 | 1654 | 1967 | 3475.45 |
| ---- | --- | ---- | --- | ---- | ---- | ---- | ------- |
| -650 | 70  | 1700 | 603 | 1902 | 1639 | 1968 | 3546.81 |
| -500 | 100 | 1720 | 608 | 1904 | 1669 | 1969 | 3620.66 |
| -400 | 140 | 1735 | 656 | 1906 | 1703 | 1970 | 3667    |
| -200 | 149 | 1750 | 692 | 1908 | 1728 | 1971 | 3770.16 |
| -10  | 175 | 1760 | 760 | 1910 | 1777 | 1972 | 3844.8  |
| 1    | 188 | 1770 | 820 | 1912 | 1788 | 1973 | 3920.25 |

| Year | Population (milion) | Year | Population (milion) | Year | Population (milion) |
| ---- | ------------------- | ---- | ------------------- | ---- | ------------------- |
| 1974 | 3995.52             | 1991 | 5406.25             | 2008 | 6811.6              |
| 1975 | 4069.43             | 1992 | 5492.69             | 2009 | 6898.31             |
| 1976 | 4142.51             | 1993 | 5577.43             | 2010 | 6985.6              |
| 1977 | 4215.77             | 1994 | 5660.73             | 2011 | 7073.13             |
| 1978 | 4289.66             | 1995 | 5743.22             | 2012 | 7161.7              |
| 1979 | 4365.58             | 1996 | 5812                | 2013 | 7250.59             |
| 1980 | 4444.01             | 1997 | 5906.48             | 2014 | 7318                |
| 1981 | 4524.63             | 1998 | 5980                | 2015 | 7426.6              |
| 1982 | 4607.98             | 1999 | 6062                | 2016 | 7492                |
| 1983 | 4691.88             | 2000 | 6148.9              | 2017 | 7599.82             |
| 1984 | 4775.84             | 2001 | 6230.75             | 2018 | 7683.79             |
| 1985 | 4861.73             | 2002 | 6312.41             | 2019 | 7743                |
| 1986 | 4950.06             | 2003 | 6393.9              | 2020 | 7840.95             |
| 1987 | 5040.98             | 2004 | 6471                | 2021 | 7909.3              |
| 1988 | 5132.29             | 2005 | 6558.18             | 2022 | 8000                |
| 1989 | 5223.7              | 2006 | 6641.42             | 2023 | 8045                |
| 1990 | 5316.18             | 2007 | 6717                |      |                     |

**REFERENCES:**

1. Abubaka I. The future of migration, human population and global health in the Anthropocene. The Lancet 2020; 396: 1133-1134. doi: 10.1016/S0140-6736(20)31523-3.

2. Tong S, Bambrick S. Sustaining planetary health in the Anthropocene. J Glob Health 2022; 12: 03068. doi:10.7189/jogh.12.03068.

3. Laublicher L. 8 billion humans: population growth, climate change and the 'Anthropocene engine'. Science. The Wire 2022: 11; 11. available at: https://science.thewire.in/environment/population-growth-climate-change-anthropocene-engine/

4. Huxley A. Brave new world. London: Penguin Random House; London: 2004. ISBN: 978-0099477464.

5. Crump T. A brief history of the Age of Steam. The power that drove industrial revolution. Robinson; London: 2007. ISBN: 978-1845295530.

6. Jonson FA. The industrial revolution in the Anthropocene. Journal of Modern History 2012; 84: 679-696; 2012. doi: 10.1086/666049.

7. Schwab K. The fourth industrial revolution. Sydney; Currency: 2017. ISBN: 978-0241300756.

8.  United Nations Department of Economic and Social Affairs, Population Division. World population prospects 2022.
https://population.un.org/w.pp/Graphs/Probabilistic/POP/TOT/900

9. T. Malthus, An Essay on the principle of population. John Murray; London: 1798. ISBN: 978-1680922585.

10. Verhulst PF. Deuxieme Memoire sur la Loi d'Accroissement de la Population. Mémoires de l'Académie Royale des Sciences, des Lettres et des Beaux-Arts de Belgique (1847) in EuDML (2022) 20, 1-32. https://eudml.org/doc/178976.

11. von Foerster H, Mora PM, Amiot LW, Doomsday: Friday 13 November, A.D. 2026. Science 1960;132: 1291-1295. doi: 10.1126/science.132.3436.1291.

12. Taagepera R. People, Skills, and resources: an interaction model for world population growth. Technol. Forecast Social Changes 1979; 13: 13-30. doi: 10.1016/0040-1625(79)90003-9.

13. Kremer A.  Population growth and technological change: one million B.C. to 1990 The Quarterly J. Econ.  1993; 108, 681-716. doi: 10.2307/2118405.

14. Golosovsky MA. Models of the world human population growth- critical analysis. 2009;  eprint arXiv:0910.30562009: 1-18.

15. Kapitza SP. On the theory of global population growth. Physics Uspekhi 2010; 53: 1287-1337. doi: 10.3367/UFNe.0180.201012g.1337.

16. Bacaër N. A short history of mathematical population dynamics. Springer; Heidelberg: 2011. ISBN 978-0-85729-114-1.

17. Barnosky AD, Ehrlich PR, Hadly EA. Avoiding collapse: grand challenges for science and society to solve by 2050. Elementa: Science of the Anthropocene 2016; 4: 000094. doi: 10.12952/journal.elementa.000094.

18. Ribeiro FL. An attempt to unify some population growth models from first principles. Revista Brasileira de Ensino de Física 2017; 39: e1311. http://dx.doi.org/10.1590/1806-9126-RBEF-2016-0118.

19. Rodrigo M, Zulkarnaen M. Mathematical models for population growth with variable carrying capacity: analytical solutions. AppliedMath 2022; 2: 466–479. doi: 10.3390/appliedmath2030027.

20. Akaev A. Phenomenological models of the global demographic dynamics and their usage for forecasting in 21st century. Appl. Math. 2022, 13, 612-649. doi: 10.4236/am.2022.137039.

21. Lueddeke GR. Global population health and well-being in the 21st century: toward new paradigms, policy, and practice. Springer, Berlin, 2015. ISBN: 978-0826127679.

22. Stokstad E. Will Malthus continue to be wrong? Science 2005; 309: 102. doi: 10.1126/science.309.5731.102.

23. Brander JA. Viewpoint: sustainability: Malthus revisited? Can J Econom. 2007; 4: 1-38. doi: 10.1111/j.1365-2966.2007.00398.x.

24. Weil DN, Wilde, J. How relevant is Malthus for economic development today?, Am Econ Rev. 2010;100: 378–382. doi: 10.1257/aer.99.2.255.

25. Kaack, LH. Katul GG. Fifty years to prove Malthus right. Proc Natl Acad Sci. 2013; 110: 4161-416. doi: 10.1073/pnas.1301246110.

26. Brown LR, Gardner GB. Beyond Malthus: the nineteen dimensions of the population challenge. New York: Earthscan; 2015. ISBN: 978-0393319064.

27. Smil V. Growth: From microorganisms to megacities. MIT Press; Cambridge MA: 2019. ISBN: 0262539683.

28. Montano B, Garcia-López MS. Malthusianism of the 21$^{st}$ century. Environ Sustain. Indicator 2020; 6: 100032. doi: 10.1016/j.indic.2020.100032.

29. Glausen JC. Thomas Malthus and global malthusianism, chapter 13, in Rahman SA, Gordy KS, Deylami SS. Globalizing political theory. Routledge; London: 2022. doi: 10.4324/9781003221708.

30. Kendall BE, Fox GA, Fujiwara M, Nogeire, TM. Demographic heterogeneity, cohort selection, and population growth. Ecology 2011; 92: 1985-1995. doi: 10.1890/11-0079.1.

31. Lima M, Berryman, AA. Positive and negative feedbacks in human population dynamics: future equilibrium or collapse? Oikos 120; 1301-1310: 2011. doi: 10.1111/j.1600-0706.2010.19112.x.

32. Cecconi F, Cencini M, Falcionia M, Vulpiani, A. Predicting the future from the past: an old problem from a modern perspective. American Journal of Physics 2012; 80: 1001-1008,. doi:10.1119/1.4746070.

33. Herrington G. Update to limits to growth: comparing the World3 model with empirical data. Journal of Industrial Ecology 2020; 25: 614-626. doi: 10.1111/jiec.13084.

34. Bystroff C. Footprints to singularity: a global population model explains late 20$^{th}$ century slow-down, and predicts peak within ten years. PLoS ONE 2021; 16: e0247214. doi: 10.1371/journal.pone.0247214.

35. Dias A, D'Hombres M, Ghisetti B, Pontarollo C, Dijkstra N. The determinants of population growth: literature review and empirical analysis. Working Papers -10, Joint Research Centre, European Commission, 2018. doi:10.2760/513062.

36. van Witteloostuijn, A, Vanderstraeten J, Slabbinck M, Dejardin J, Hermans J, Coreynen W. From explanation of the past to prediction of the future: a comparative and predictive research design in the social sciences. J Soc Sci Humanities 2022 ;6: 100269. doi: 10.1016/j.ssaho.2022.100269.

37. McFarlane I. State world population report 2023. 8 billion lives, infinite possibilities – the case for rights and choices (UNFPA, New York, 2023). www.unfpa.org/swp2023.

38. Lehman C, Loberg S, Wilson M, Girham E. Ecology of the Anthropocene signals hope for consciously managing the planetary ecosystem. Proc Natl Acad Sci. 2021; 118: e2024150118. doi:10.1073/pnas.2024150118.

39. Pearl R. The growth of populations. Quter Rev Biol. 1927; 2: 532–548. doi: 10.1126/science.66.1702.x.t.

40. Pearl, R, Reed L. On the rate of growth of the population of the United States since 1790 and its mathematical representation. Proc Natl Acad Sci 1920; 6: 275-88. doi: 10.1073/pnas.6.6.275.

41. Harper WL. Isaac Newton's scientific method: turning data into evidence about gravity and cosmology. Oxford Univ. Press., Oxford, 2012. ISBN: 978-0199570409.

42. Rzoska AA, Drozd-Rzoska A. The story about one island and four cities. The socio-economic soft matter model - based report. Proc. 8[th] Socratic Lectures 2023; 8: 131-147. https://doi.org/10.55295/PSL.2023.

43. Murray D. Mathematical biology: i. an introduction. Springer-Verlag; Berlin: 2002. ISBN: 978-0387952239.

44. Stanescu D, Chen-Charpentier BM. Random coefficient differential equation models for bacterial growth. Mathematical and Computer Modelling 2009; 50: 885-895, doi: 10.1016/j.mcm.2009.05.017.

45. Morales-Eros AJ, Reyes-Reyes J, Astorga-Zaragoza CM, Osorio-Gordillo GL, García-Beltrán CD, Madrigal-Espinosa G. Growth modeling approach with the Verhulst coexistence dynamic properties for regulation purposes. Theory Biosci. 2023; 142: 221–234. doi: 10.1007/s12064-023-00397-x.

46. Gogoi UN, Saikia P, Mahanta DJ. A Review of the growth models in biological sciences. AIP Conference Proceedings 2022; 451: 020070. doi: 10.1063/5.0095187

47. Peleg JM, Corradini MG, Normand MD. The logistic (Verhulst) model for sigmoid. Food Res Internat. 2007; 40: 808-818. doi: 10.1016/j.foodres.2007.01.012.

48. Mahajan TS, Pandey OP. Reformulation of Malthus-Verhulst equation for black gram seeds pretreated with magnetic field. International Agrophys. 2011; 25: 355-359. ISSN:0236-8722.

49. Vandamme LKJ, Rocha PRF, Analysis and simulation of epidemic covid-19 curves with the Verhulst model applied to statistical inhomogeneous age groups. Appl Sci. 2021; 11: 4159. doi: 10.3390/app11094159.

50. Morales-Erosa AJ, Reyes-Reyes J, Astorga-Zaragoza CM, Osorio-Gordillo GL, García-Beltrán CD, Madrigal-Espinosa G. Growth modeling approach with the Verhulst coexistence dynamic properties for regulation purposes. Theory Biosci. 2023; 142: 221–234. doi: 10.1007/s12064-023-00397-x,

51. Iskender C. Mathematical study of the Verhulst and Gompertz growth functions and their contemporary applications. Ekoist: Journal of Econometrics and Statistics 2021; 34: 73-102. doi: 10.26650/ekoist.2021.34.876749.

52. Kwasnicki, W. Logistic growth of the global economy and competitiveness of nations, Technol. Forecast Social Change 2013; 80: 50-76,. doi: 10.1016/j.techfore.2012.07.007.

53. Gleriaa I, Da Silvab S, Brenig L, Rocha Filho TM, Figueiredo A. A Modified Verhulst-Solow model for long-term population and economic growths. 2023. doi: 10.48550/arXiv.2308.08315.

54. Jena A, Mishra BS. Ordering kinetics and steady state of Malthusian flock. Physics of Fluids 2023; 35: 107138. doi: 10.1063/5.0167463.

55. de Pasquale F, Tartaglia P, Tombesi P. Transient behavior of nonequilibrium phase transitions. Phys Lett A 1980; 78: 129-132. doi:10.1016/0375-9601(80)90675-1.

56. Dekker H. On the critical point of a Malthus–Verhulst process. J Chem Phys. 1980; 72: 189–191. doi: 10.1063/1.438901.

57. Agenbroad LD. Pygmy (dwarf) mammoths of the channel islands of california. Mammoth Site of Hot Springs: Dallas, TX, USA:1998. ISBN: 978-0962475085

58. Kennett DJ, Kennett JP, West GJ, Erlandson JM, Johnson JR, Hendy IL, West A, Culleton B., Jones TL, Stafford T.W. Wildfire and abrupt ecosystem disruption on California's Northern Channel Islands at the Ållerød–Younger Dryas boundary (13.0–12.9ka). Quarter Sci Rev. 2008; 27: 2530-2545. doi: 10.1016/j.quascirev.2008.09.006.

59. Newfield TP. Mysterious and mortiferous clouds: the climate cooling and disease burden of late antiquity. Late Antique Archaeology 2016; 12: 89-115. doi: 10.1163/22134522-12340068.

60. Heather P. Empires and barbarians: migration, development and the birth of Europe. Pan Macmillan. London: 2010. ISBN 978-0-330-54021-6.

61. Gerard Labuda: Słowiańszczyna starożytna i wczesnośredniowieczna (in polish, *title in english: Ancient and early medieval Slavic region*). Poznań (Poland); WPTPN: 2003. ISBN: 83-7063-381-1.

62. Profantova N. Cultural discontinuity and the migration hypothesis. The 6th-century Slavic migration in the light of new archaeological finds from Bohemia. ACE Conference Brussels: The very beginning of Europe? Early-medieval migration and colonisation. 2012), pp. 255-264.

63. Stolarek I., Zenczak M, Handschuh L. et al. Genetic history of East-Central Europe in the first millennium CE. Genome Biol. 2023; 24: 173. doi:10.1186/s13059-023-03013-9.

64. Rzoska AA. Econo- and socio- physics based remarks on the economical growth of the World. Turk. Econ. Rev. 3, 82-90, 2016. ISSN: 2149-0414.

65. Volterra, V. Variations and fluctuations of the number of individuals in animal species living together. J Con. Int Explor Mer. 1928; 3: 3-51(. doi:10.1093/icesjms/3.1.3.

66. A Drozd-Rzoska, Universal behavior of the apparent fragility in ultraslow glass forming systems. Scientific Reports 2019; 9: 6816. doi: 10.1038/s41598-019-42927-y.

67. Drozd-Rzoska, A. Activation volume in superpressed glass-formers. Scientific Reports 2019; 9: 13787. doi: 10.1038/s41598-019-49848-w.

68. Drozd-Rzoska A. Pressure-related universal previtreous behavior of the time and apparent fragility. Front Mater. 2019; 6: 103. doi: 10.3389/fmats.2019.00103.

69. Drozd-Rzoska A, Rzoska SJ, Starzonek S. New scaling paradigm for dynamics in glass-forming systems. Prog. Mater. Sci. 2023; 134: 101074. doi: 10.1016/j.pmatsci.2023.101074.

70. United States Census Bureau: https://www.census.gov/population

71. Taagepera R, Nemčok M. World population growth over millennia: Ancient and present phases with a temporary halt in-between. The Anthropocene Review 2023, 0(0). doi: 10.1177/20530196231172.

72. https://populationmatters.org/the-facts-numbers

73. World bank 2022 – 1960: https://www.macrotrends.net/countries/WLD/world/population

74. https://www.statista.com/statistics/1006502/global-population-ten-thousand-bc-to-2050/

75. https://en.wikipedia.org/wiki/Estimates_of_historical_world_population

76. Federico G, Junguito AT,. How many people on earth? World population 1800-1938. The Center for Economic Policy Research (CEPR). VOX EU, Brussels, 2023. https://cepr.org/voxey/columns/how-many-people-earth-world-population-1800-1938.

77. McEvedy C, Jones R. Atlas of World population history, facts on file. Puffin: New York: 1978. ISBN: 978-0871964021.

78. Glass B. The Anthropocene epoch. When humans changed the World. Houston TX; DBG Publishing: 2021. ISBN: 978-0578995304.

79. Garnsey P. The Roman Empire: economy, society and culture. Oakland; University of California Press: 2014. ISBN: 978-0520285989.

80. Harper K. The fate of Rome: climate, disease, and an end of the Empire. Princeton; Princeton University Press: 2017. ISBN: 978-0691166834.

81. Bowman A, Wilson A. Quantifying the Roman Economy: Methods and Problems. Oxford: Oxford University Press: 2009. ISBN: 978-0199562596.

82. Wickham C. Medieval Europe. New Haven; Yale University Press: 2016. ISBN: 978-0300208344.

83. Watts F. The Emperor Charlemagne. London; Franklin Watts: 1986. ISBN: 978-0531150047.